\documentclass[aps,graphicx,twocolumn]{revtex4}
\usepackage{amsmath}
\usepackage{amscd}
\usepackage{graphicx}
\usepackage{CJK}
\usepackage{multirow}
\begin{document}

\begin{CJK*}{GBK}{song}
\title{Control power in perfect controlled teleportation via partially entangled channels}
\author{ Xi-Han Li$^{1,2}$ and Shohini Ghose$^{1,3}$\footnote{
Email address: sghose@wlu.ca}}
\address{$^1$Department of Physics and Computer Science, Wilfrid Laurier University, Waterloo, Canada \\$^2$Department of Physics, Chongqing University,
Chongqing, China\\
$^3$Institute for Quantum Computing, University of Waterloo, Canada}

\date{\today }
\begin{abstract}
We analyze and evaluate perfect controlled teleportation via three-qubit entangled channels from the point of view of the controller. The key idea in controlled teleportation is that the teleportation is performed only with the participation of the controller. We calculate a quantitative measure of the controller's power and establish a lower bound on the control power required for controlled teleportation. We show that the maximally entangled GHZ state is a suitable channel for controlled teleportation of arbitrary single qubits - the controller's power meets the bound and the teleportation fidelity without the controller's permission is no better than the fidelity of a classical channel.  We also construct partially entangled channels that exceed the bound for controlled teleportation of a restricted set of states called the equatorial states. We calculate the minimum entanglement required in these channels to exceed the bound.  Moreover, we find that in these restricted controlled teleportation schemes, the partially entangled channels can outperform maximally entangled channels with respect to the controller's power. Our results provide a new perspective on controlled teleportation schemes and are of practical interest since we propose useful partially entangled channels.
\end{abstract}

\maketitle

\section{Introduction}
Entanglement is a phenomenon unique to the quantum world and is an important resource for quantum information processing. 
It is widely used in quantum information processing, such as quantum key distribution \cite{qkd1,qkd2,qkd3}, 
quantum secret sharing \cite{qss1,qss2,qss3}, 
quantum dense coding \cite{dense1,dense2}, quantum secure direct communication \cite{qsdc1,qsdc2} and quantum computation\cite{qc1,qc2}. One of the most intriguing uses of entanglement is for  quantum teleportation \cite{tele}. An arbitrary quantum state can be recovered in a remote location with the aid of a maximally entangled Einstein-Podolsky-Rosen (EPR) pair and two bits of classical information.
Quantum teleportation has been widely studied theoretically and experimentally in the past twenty years \cite{t1,t2,t3,t4}.
Standard quantum teleportation involves only two parties. The sender and the receiver share 
a maximal entanglement in advance. After the sender's Bell state measurement on the state to be teleported and one of the entangled 
particle, the receiver can obtain the original state with proper unitary operation according to the sender's measurement result, i.e.,
 the state is teleported from the sender to the receiver. A variant of quantum teleportation called controlled teleportation (CT) was 
 first proposed in 1998 \cite{ct}. In this scheme, the teleportation procedure is controlled by a controller, such that the arbitrary 
 quantum state can be teleported from sender to receiver only with the participation of the controller \cite{qss}. 
 The protocol described in \cite{ct} utilized the maximally entangled 3-qubit  Greenberger-Horne-Zeilinger (GHZ) state 
 as the quantum channel for CT of a single qubit (we call it the GHZ scheme in the following).  Controlled teleportation 
 is useful in various contexts in quantum communication, including in quantum networks and cryptographic conferences 
 \cite{cta1,cta2,cta3,cta4}. Following the GHZ scheme, a number of controlled teleportation schemes have been proposed 
 thus far \cite{ct1,ct2,ct3,ct4,zhou,gao,Kumar}. From an intuitive point of view, one would think that maximally entangled states are required as a high quality quantum resource in controlled teleportation for optimal performance, just like in other quantum information processing tasks, although it can be a technological challenge to prepare and share maximal entanglement in practical experiments.

 In 2008 Gao \emph{et al.} ~\cite{gao} found that certain partially entangled states called maximal slice (MS) states~\cite{ms} 
 can also be used for controlled teleportation. The CT scheme employing the MS states has 100\% success probability and fidelity 
 of teleportation, which is the same as the GHZ scheme \cite{ct}. However, we show in this paper that these schemes are different 
 from the controller's point of view, i.e., the controller's power is different. Although a lot of work has been devoted to studying 
 controlled teleportation, very little has been discussed about the controller's measurable authority. In a controlled teleportation 
 scheme, it is important and necessary to ensure the controller's authority while retaining the success probability and fidelity of 
 teleportation.
We define a measure of the controller's power based on the non-conditioned fidelity (NCF) - the teleportation fidelity achievable without the controller's permission and participation. This non-conditioned fidelity of teleportation must be minimized in order to maximize the controller's power. We show that in the GHZ scheme, the teleportation fidelity that can be achieved without the controller's permission is no better than the fidelity using a classical channel \cite{ct}. Thus the maximally entangled GHZ  state is a suitable channel for CT  of arbitrary single qubits as it ensures the controller's authority -  the teleportation fidelity can only be greater than the classical limit with the controller's participation. On the other hand, when MS states are used, then the teleportation fidelity achieved  without the controller's permission can be greater than the classical limit and hence these states are not suitable channels for CT of arbitrary single qubit states.
However, we  find that  the MS states and other similar partially entangled states are good channels for controlled teleportation of certain restricted sets of input states - the equatorial states \cite{equatorial1,equatorial2}. For these restricted input states, the controller's power is preserved, and the teleportation fidelity cannot be greater than the classical limit without the controller's permission. We show that these partially entangled channels can even outperform the maximally entangled states in maximizing the controller's power. We calculate the minimum degree of entanglement required in the partially entangled quantum channel that will ensure the controller's authority in CT schemes. Our work provides  a new perspective on quantum controlled teleportation and on the properties of three-qubit entanglement. Our results are of relevance for designing practical implementations of teleportation using non-maximal entanglement.

\section{controlled teleportation via Maximal Slice states}
The MS states can be written as \cite{gao,ms}
\begin{eqnarray}
\vert MS \rangle_{123}&=&\frac{1}{\sqrt{2}}(\vert 000 \rangle +c\vert 111\rangle +d\vert 011\rangle)_{123},
\end{eqnarray}
where $c$ and $d$ are assumed to be real and $c^2+d^2=1$.
When $c=0$, the resulting state is a product state of the first qubit with a maximally entangled Bell pair of qubits 2 and 3.  When $c=1$, the resulting state is  the maximally entangled GHZ state. For all other values of $c$, the three qubits are partially entangled. The 3-qubit entanglement as measured by the 3-tangle~\cite{tangle} is $c^2$. The MS states have been shown to have interesting entanglement and nonlocality properties due to their inherent symmetries~\cite{ms, nonlocality}. Furthermore, Gao \emph{et al.} mathematically showed that all states, equivalent under local unitaries to the MS states, can be used for performing perfect, deterministic controlled teleportation. Here, we present a simple way to understand the Gao result.

Suppose the  arbitrary state to be teleported is
\begin{eqnarray}
\vert \varphi\rangle_t=k_0\vert 0\rangle_t+k_1\vert 1\rangle_t, (\vert k_0\vert^2+\vert k_1\vert^2=1).
\end{eqnarray}
The three qubits of the MS state are distributed to the controller Charlie, who gets qubit 1, the sender Alice, who gets qubit 2 and the receiver Bob, who gets qubit 3.
 The MS state can be rewritten as
\begin{eqnarray}
\vert MS \rangle_{123}&=&\frac{1}{2}[(1+d)\vert 0\rangle+c\vert 1\rangle]_1\otimes\vert \Phi^+\rangle_{23}\nonumber\\
&&+\frac{1}{2}[(1-d)\vert 0\rangle-c\vert 1\rangle]_1\otimes\vert \Phi^-\rangle_{23},
\end{eqnarray}
where $\vert \Phi^{\pm}\rangle=\frac{1}{\sqrt{2}}(\vert 00\rangle\pm\vert 11\rangle)$ are the two Bell states. This structure of the MS state as a superposition of Bell states makes it easy to see why it can be used for perfect controlled teleportation; if Charlie measures his qubit 1 in the following orthogonal basis,
\begin{eqnarray}
\vert x_+\rangle=\frac{1}{\sqrt{(1+d)^2+c^2}}[(1+d)\vert 0\rangle+c\vert 1\rangle_1],\nonumber\\
\vert x_-\rangle=\frac{1}{\sqrt{(1-d)^2+c^2}}[(1-d)\vert 0\rangle-c\vert 1\rangle_1],
\end{eqnarray}
then Alice and Bob will always be left with one of the two maximally entangled Bell states $\vert \Phi^{\pm}\rangle$ depending on Charlie's measurement outcome. If Charlie broadcasts his measurement outcome to Alice and Bob, then they will know
which Bell state they are sharing and can then use it for teleportation in the standard way. The success probability and the fidelity of this scheme are both 100\%. It thus appears that the partially entangled MS state can implement controlled teleportation as well as the maximally entangled GHZ state. What is even more surprising is that perfect CT seems to be possible with MS states regardless of the degree of entanglement.
However,  a more careful analysis of the controller's power shows that the MS states have some limitations in their use for controlled teleportation.

Let us compute the non-conditioned fidelity (NCF), the fidelity of the teleportation without Charlie's collaboration. 
It is necessary to point out that the NCF is calculated with the sender's participation. The state of the joint quantum system composed of $\vert \varphi\rangle_t$ and $\vert MS \rangle_{123}$ can be rewritten in terms of the Bell basis as
\begin{eqnarray}
&&\vert \varphi\rangle_t\otimes \vert MS \rangle_{123}\nonumber\\&=&\frac{1}{2}\vert \Phi^+\rangle_{t2}(k_0 \vert 00 \rangle +k_1c\vert 11\rangle +k_1d\vert 01\rangle)_{13}\nonumber\\
&&+\frac{1}{2}\vert \Phi^-\rangle_{t2}(k_0 \vert 00 \rangle -k_1c\vert 11\rangle -k_1d\vert 01\rangle)_{13}\nonumber\\
&&+\frac{1}{2}\vert \Psi^+\rangle_{t2}(k_1 \vert 00 \rangle +k_0c\vert 11\rangle +k_0d\vert 01\rangle)_{13}\nonumber\\
&&-\frac{1}{2}\vert \Psi^-\rangle_{t2}(k_1 \vert 00 \rangle -k_0c\vert 11\rangle -k_0d\vert 01\rangle)_{13}.
\end{eqnarray}

For Alice's different Bell measurements of qubits 2 and $t$,
Bob and Charlie's qubits collapsed onto the corresponding states as shown above. Then  the density matrix describing Bob's qubit 3,  while tracing over qubit 1 is
\begin{eqnarray}
\rho_3= tr_1({\vert \psi \rangle_{13}\langle \psi \vert }).
\end{eqnarray}

The density matrix $\rho_3$ can be transformed into the same one for all of Alice's different outcomes with proper unitary operations performed by Bob. Then the NCF can be computed by
\begin{eqnarray}
f=\langle \varphi \vert \rho_3 \vert \varphi\rangle,
\end{eqnarray}
where $\vert \varphi\rangle$ is the desired state to be teleported.
The non-conditioned fidelity of teleportation using the MS state is
\begin{eqnarray}
f_{MS}&=&\vert k_0 \vert^4+\vert k_1 \vert^4+2 \vert d\vert \vert k_0 \vert^2\vert k_1 \vert^2,
\end{eqnarray}
which depends on the state to be teleported. In order to calculate the average fidelity over all input
states which are assumed appear equally often, the parameters are rewrite in polar coordinates as
\begin{eqnarray}
k_0=\cos\theta, k_1=e^{i\phi}\sin\theta.
\end{eqnarray}
Then the average fidelity can be computed by
\begin{eqnarray}
\bar{f}_{MS}=\frac{1}{4\pi}\int^{2\pi}_{0}d\phi\int_0^{\pi}f_{MS}\sin\theta d\theta
\end{eqnarray}
and we get
\begin{eqnarray}
\bar{f}_{MS}&=&\frac{2}{3}+\frac{\vert d\vert}{3}.\label{fms}
\end{eqnarray}
From this expression, we see that when $d=0$, the result is consistent with that of the GHZ scheme \cite{ct}. And when $d=1$, Bob can recover the input state perfectly without Charlie's help since the original channel is a product state between Charlie's qubit and the rest of the system. For a general MS state, Bob's average non-conditioned fidelity is always larger than $2/3$.

The aim of controlled teleportation is to teleport an arbitrary quantum state from the sender to the receiver,
but only with the permission of the controller.  Therefore, the NCF which measures the fidelity of the teleportation
without the controller's participation (i.e, permission), should be minimized in order to maximize the controller's authority.
We can thus define Charlie's control power as
 \begin{eqnarray}
 C=1-f
  \end{eqnarray}
 In this case, the average control power is $\bar{C}=1-\bar{f}.$
As shown in Ref.\cite{2/3,2/3+}, $2/3$ is the optimal value of the fidelity for estimating the quantum state with only one sample. It is also called the classical fidelity since it is the maximum possible fidelity when two parties communicate with each other only through a classical channel\cite{ct,2/3,2/3+}. A controlled teleportation scheme should ensure that the receiver cannot achieve better than the classical fidelity without the controller's permission.
So $\bar C$ should be no less than $1/3$.

From Eq. (\ref{fms}), we find that the fidelity of the GHZ scheme $d=0$ is exactly the classical limit. And for $d=1$, Bob can recover the teleported state without the help of controller, which makes the controller powerless. For all other MS states, $\bar{C}_{MS}$ is always less than $1/3$. In other words, the teleportation fidelity can always exceed the classical limits without Charlie's help.  Thus the general MS states are unsuitable for controlled teleportation of arbitrary states since the controller's power is less than the classical limit. However, as we show in the following section, MS states and other similar partially entangled channels are suitable for controlled teleportation of certain subsets of states known as equatorial states.

\section{Perfect controlled teleportation of equatorial states via partially entangled channels}
The equatorial states are states whose Bloch vector is restricted to the intersection
of the $x-z$ ($x-y$, $y-z$) plane with the Bloch sphere \cite{equatorial1,equatorial2}. The $y (z, x)$ component of the Bloch vector is zero for these states. For simplicity, we call these three kinds of states the $x-z$ state, the $x-y$ state and the $y-z$ state, respectively. These states can be written as
\begin{eqnarray}
\vert \varphi_{x-z}\rangle_t&=&\cos\frac{\theta}{2}\vert 0\rangle_t +\sin \frac{\theta}{2}\vert 1\rangle_t,\\
\vert \varphi_{x-y}\rangle_t&=&\frac{1}{\sqrt{2}}(\vert 0\rangle_t +e^{i\phi}\vert 1\rangle_t),\\
\vert \varphi_{y-z}\rangle_t&=&\cos\frac{\theta}{2}\vert 0\rangle_t +i\sin \frac{\theta}{2}\vert 1\rangle_t.
\end{eqnarray}
which are specific subclasses of the arbitrary single-qubit state $\vert \varphi\rangle_t=\cos\frac{\theta}{2}\vert 0\rangle_t +e^{i\phi}\sin \frac{\theta}{2}\vert 1\rangle_t$.

Our goal is to construct quantum channels for teleporting these states such that Charlie's control power exceeds the classical limit of 1/3. Like the MS states, we start by constructing a partially entangled superposition of Bell states  that is useful for perfect controlled teleportation:\begin{eqnarray}
\vert \Theta \rangle_{123}= a\vert 0\rangle_1 \vert \Phi^+\rangle_{23}+b\vert 1\rangle_1 \sigma_{k3}\vert \Phi^+\rangle_{23}.
\end{eqnarray}
Here $a^2+b^2=1$ and the qubits 1, 2, 3 are distributed to Charlie, Alice and Bob, respectively. $\sigma_{k3}(k=x,y,z)$ are the three Pauli operators acting on qubit 3:
\begin{eqnarray}
\sigma_x=\left( {\begin{array}{*{20}{c}}
0&1\\
1&0
\end{array}} \right),\sigma_y=\left( {\begin{array}{*{20}{c}}
0&-1\\
1&0
\end{array}} \right),\sigma_z=\left( {\begin{array}{*{20}{c}}
1&0\\
0&-1
\end{array}} \right).
\end{eqnarray}
The state in Eq.~(14) can always be used for perfect teleportation since the state shared by Alice and Bob will be a Bell state after Charlie's measurement of qubit 1 in the $\vert 0\rangle,\vert 1\rangle$ basis. Given Charlie's measurement results, Alice and Bob know which Bell state they share and can use it for teleportation in the usual way.

We now consider the situation where Alice and Bob want to proceed with the teleportation without Charlie's permission - i.e. - without his participation. In that case, Charlie does not measure his qubit. Alice performs a Bell state measurement on her qubit 2 and qubit $t$ which is the qubit to be teleported. The remaining joint state of Charlie's qubit 1 and Bob's qubit 3 can always be transformed via local operations to
\begin{eqnarray}
\vert \psi \rangle_{13}=a\vert 0\rangle_1 \otimes \vert \varphi_{j} \rangle_3 + b\vert 1\rangle_1 \otimes\sigma_{k3}\vert \varphi_{j} \rangle_3.
\end{eqnarray}
Here $j$ represent the three possible sets of input states.
The reduced density matrix of Bob's qubit is thus
\begin{eqnarray}
\rho_3=a^2 \vert \varphi_{j} \rangle_3 \langle \varphi_{j} \vert+ b^2 \sigma_{k3}\vert \varphi_{j} \rangle_3 \langle \varphi_{j} \vert \sigma^\dag_{k3}.
\end{eqnarray}
The non-conditioned fidelity $\langle \varphi_{j} \vert \rho_3\vert \varphi_{j} \rangle$  is then
\begin{eqnarray}
f&=&a^2 \vert \langle \varphi_{j} \vert \varphi_{j} \rangle \vert^2 +b^2 \vert \langle \varphi_{j} \vert\sigma_{k3}\vert \varphi_{j} \rangle \vert^2\nonumber\\
&=&a^2+b^2 \vert \langle \varphi_{j} \vert\sigma_{k3}\vert \varphi_{j} \rangle \vert^2.
\end{eqnarray}
To maximize Charlie's control power, the non-conditioned fidelity of teleportation achievable without the controller's permission must be minimized. From the above expression, we see that the minimum value of $f$ can be obtained when $\vert \langle \varphi_{j} \vert\sigma_{k3}\vert \varphi_{j} \rangle \vert^2=0$.
For our three sets of equatorial input states, we can find corresponding Pauli operators $\sigma_{k3}$ to get the minimum
\begin{eqnarray}
\vert \langle \varphi_{x-z} \vert\sigma_{y}\vert \varphi_{x-z} \rangle \vert^2 =0,\\
\vert \langle \varphi_{x-y} \vert\sigma_{z}\vert \varphi_{x-y} \rangle \vert^2 =0,\\
\vert \langle \varphi_{y-z} \vert\sigma_{x}\vert \varphi_{y-z} \rangle \vert^2 =0.
\end{eqnarray}
From the symmetry of the parameters $a$ and $b$, we get $f_{min}=max(a^2,b^2)$, which is independent of the parameters of the input states. We thus do not need to average over all input  equatorial states. The corresponding partially entangled channels for the equatorial states can be obtained by substituting the Pauli operators $\sigma_{k3}$ from Eq. (19 -21) into Eq. (14). For the $x-z$ state with real parameter, the resulting quantum channel is thus
\begin{eqnarray}
\vert \Theta_{x-z}\rangle_{123}= a\vert 0\rangle_1 \vert \Phi^+\rangle_{23}+b\vert 1\rangle_1 \vert \Psi^-\rangle_{23}.
\end{eqnarray}
This state is also known as a three-qubit tetrahedral state \cite{ms,mt}.
For the $x-y$ state, the quantum channel is
\begin{eqnarray}
\vert \Theta_{x-y}\rangle_{123}= a\vert 0\rangle_1 \vert \Phi^+\rangle_{23}+b\vert 1\rangle_1 \vert \Phi^-\rangle_{23}.
\end{eqnarray}
This is the MS state.
Finally, for the $y-z$ state with fixed relative phase, the quantum channel is
\begin{eqnarray}
\vert \Theta_{y-z}\rangle_{123}= a\vert 0\rangle_1 \vert \Phi^+\rangle_{23}+b\vert 1\rangle_1 \vert \Psi^+\rangle_{23}.
\end{eqnarray}

To sum up, we have found partially entangled quantum channels for CT  of equatorial states  such that the non-conditioned fidelity is minimized.
The next question is whether this minimum non-conditioned fidelity results in a control power that is equal to or better than the classical limit of 1/3.
For all the three cases, the control power is
\begin{eqnarray}
C=1-f_{min}=1-max(a^2,b^2).
\end{eqnarray}
Thus $C\geq1/3$ when
\begin{eqnarray}
1/3\leq a^2\leq2/3.
\end{eqnarray}
For these values, the fidelity of teleportation without the controller's permission is no better than the classical limit of 2/3, and hence the control power is no less than the classical limit of 1/3.
If we use the three-tangle \cite{tangle} to quantify the degree of entanglement, then $\tau_{channel}=4a^2b^2$ should be no less than $8/9$ to ensure a control power equal to or greater than $1/3$.

Generally speaking, the minimum fidelity of teleportation is $1/2$, which can be obtained by merely selecting a state at random. 
Accordingly, $C_{max}=1-f_{min} =1/2$ is the maximum control power in the controlled teleportation scheme. 
In above schemes, the maximal control power can be obtained by using the corresponding maximally entangled channel, $a=b$. 
For $a \neq b$ in the partially entangled channels introduced above, the controller's power is determined by the parameters $a, b$ of the channel.

\section{Discussion and summary}
In this paper, we have investigated perfect controlled teleportation schemes via non-maximally entangled channels. The important idea in controlled teleportation is that the teleportation is performed only with the permission of the controller. We have therefore analyzed CT schemes by defining a measure of the  controller's power based on the teleportation fidelity achievable without the controller's permission. The controller's power should be no less than a minimum of 1/3 to ensure that the fidelity of teleportation without the controller's permission is no better than the fidelity using a classical channel. Based on this measure, we showed that for CT of arbitrary single-qubit unknown states, the maximum control power is equal to the classical limit of 1/3 and can be achieved via a maximally entangled GHZ channel. However, the partially entangled MS states are  unsuitable for controlled teleportation of arbitrary states from the controller's point of view, since the controller's power is less than the classical limit of 1/3.  Better control power can be achieved if we focus on restricted sets of input states to be teleported, i.e., the equatorial states which are popular in quantum communication. We constructed suitable partially entangled quantum channels that can achieve controlled teleportation of equatorial states with a control power no less than the classical limit of 1/3.  The entanglement of these channels as measured by the 3-tangle must be no less than 8/9 for the controller's power to beat the classical limit.  The control power is independent of the input equatorial states.

What if we used a mismatched quantum channel to teleport equatorial states? For example, suppose we use $\Theta_{i}$ to teleport $\vert \varphi_{j}\rangle (i,j\in\{x-y,y-z,x-z\}$ and $i\neq j)$. Then the non-conditioned fidelity will be
\begin{eqnarray}
f=a^2+b^2\vert \langle \varphi_{j}\vert \sigma_k \vert \varphi_{j}\rangle \vert ^2,
\end{eqnarray}
where $k=\{x,y,z\}$ corresponding to $i=\{y-z,x-z,x-y\}$, respectively. It is easy to calculate the average value of these three Pauli operators over the restricted sets of equatorial states. The results show that the non-conditioned fidelity is always larger than the classical limit, indicating inadequate  control power. Even for the maximally entangled case of $a=b$, the maximum control power is just the classical limit. In comparison, the correctly matched quantum channel (i.e., $\Theta_{j}$ to teleport $\vert \varphi_{j}\rangle (j\in\{x-y,y-z,x-z\}$) will result in a better-than classical control power when $a^2 > 1/3$. Thus in this case the correctly matched partially entangled quantum channel can outperform the mismatched quantum channel even when the mismatched channel is maximally entangled.
Therefore,  it is important that the correct quantum channel  is shared in advance between the three parties. Note that the three partially entangled channels in Eqs. (22) - (24)  are unitarily equivalent to each other.
However, the transformation cannot be realized by only one party. Therefore, the sender and the controller can prevent the receiver from rotating the channel to get a higher NCF. It is interesting that unitarily equivalent states have different performance in controlled teleportation tasks. This can be explained by the fact that unitary operations change the fidelity for our restricted sets of states.

In summary, we have evaluated the use of partially entangled quantum channels for CT and shown their advantages for
teleporting restricted sets of equatorial quantum states, which are commonly used in quantum communication schemes. 
 High control power can be obtained while retaining unit success probability and state fidelity.
Compared to the maximally entangled states, the option of using partially entangled states is attractive because of  the practical challenges of generating and maintaining maximal entanglement. Preparation of partially entangled states may be more realistic in physical systems, and could allow for more robust and flexible schemes. In real systems, the prepared states are often mixed states and in future work we plan to investigate the use of mixed states for controlled teleportation.

\section*{Acknowledgement}
XL is supported by the National Natural Science
Foundation of China under Grant No. 11004258 and the Fundamental Research Funds for the Central Universities under
Grant No.CQDXWL-2012-014. SG acknowledges support from the Ontario Ministry of Research and Innovation and the Natural Sciences and Engineering Research Council of Canada.

\end{CJK*}


\begin{thebibliography}{99}

\bibitem{qkd1} A. K. Ekert, Phys. Rev. Lett. \textbf{67}, 661 (1991).

\bibitem{qkd2} C. H. Bennett, G. Brassard, and N. D. Mermin, Phys. Rev. Lett. \textbf{68}, 557 (1992).

\bibitem{qkd3} F. G. Deng and G. L. Long, Phys. Rev. A \textbf{68}, 042315 (2003)

\bibitem{qss1} M. Hillery, V. B$\check{u}$zek, and A. Berthiaume, Phys. Rev. A \textbf{59}, 1829 (1999)

\bibitem{qss2} A. Karlsson, M. Koashi, and N. Imoto, Phys. Rev. A \textbf{59}, 162 (1999)

\bibitem{qss3} L. Xiao, G. L. Long, F. G. Deng, and J. W. Pan, Phys. Rev. A \textbf{69}, 052307 (2004)

\bibitem{dense1} C. H. Bennett and S. J. Wiesner, Phys. Rev. Lett. \textbf{69}, 2881 (1992).

\bibitem{dense2} C. Wang, F. G. Deng, Y. S. Li, X. S. Liu, and G. L. Long, Phys. Rev. A \textbf{71}, 044305 (2005).

\bibitem{qsdc1} G. L. Long and X. S. Liu, Phys. Rev. A \textbf{65} 032302 (2002). 

\bibitem{qsdc2} F. G. Deng, G. L. Long, and X. S. Liu, Phys. Rev. A \textbf{68}, 042317 (2003).

\bibitem{qc1} M. A. Nielsen and I. L.Chuang, Quantum Computation and Quantum Information
(Cambridge University Press, Cambridge, England, 2000).

\bibitem{qc2} B. C. Ren and F. G. Deng, Sci. Rep. \textbf{4}, 4623 (2014).

\bibitem{tele} C. H. Bennett, G. Brassard, C. Crepeau, R. Jozsa, A. Peres, and W. K. Wootters, Phys. Rev. Lett. \textbf{70} 1895(1993).

\bibitem{t1} D. Bouwmeester, J. W. Pan, K. Mattle, M. Eibl, H. Weinfurter, and A. Zeilinger, Nature \textbf{390}, 575 (1997).

\bibitem{t2} A. Furusawa, J. L. Sorensen, S. L. Braunstein, C. A. Fuchs, H. J. Kimble, and E. S. Polzik, Science \textbf{282},
706 (1998).

\bibitem{t3} D. Boschi, S. Branca, F. DeMartini, L. Hardy, and S. Popescu, Phys. Rev. Lett. \textbf{80}, 1121
(1998).

\bibitem{t4} Y. H. Kim, S. P. Kulik, and Y. Shih, Phys. Rev. Lett. \textbf{86}, 1370 (2001).

\bibitem{ct} A. Karlsson and M. Bourennane, Phys. Rev. A \textbf{58}, 4394 (1998).

\bibitem{qss} M. Hillery, V. Bu$\breve{z}$ek, and A. Berthiaume, Phys. Rev. A \textbf{59}, 1829 (1999).

\bibitem{cta1} E. Biham, B. Huttner, and T. Mor, Phys. Rev. A \textbf{54}, 2651 (1996).

\bibitem{cta2} P. D. Townsend, Nature, 385, \textbf{47} (1997) .

\bibitem{cta3} S. Bose, V. Vedral, and P. L. Knight, Phys. Rev. A \textbf{57}, 822 (1998).

\bibitem{cta4} B. Aoun and M. Tarifi, e-print quant-ph/0401076 (2004).

\bibitem{ct1} C. P. Yang, S. I. Chu, and S. Han, Phys. Rev. A \textbf{70}, 022329 (2004).

\bibitem{ct2} Y. Yeo, e-print quant-ph/0323030v1 (2003).

\bibitem{ct3} F. G. Deng, C. Y. Li, Y. S. Li, H. Y. Zhou, and Y. Wang, Phys. Rev. A \textbf{72},022338 (2005).

\bibitem{ct4} X. H. Li, F. G. Deng, and H. Y. Zhou, Chin. Phys. Lett. \textbf{24}, 1151 (2007).

\bibitem{zhou} P. Zhou, X. H. Li, F. G. Deng, J. Phys. A: Math. Theor. \textbf{40}, 13121 (2007). 

\bibitem{gao} T. Gao, F. L. Yan, and Y. C. Li, Euro. Phys. Lett. \textbf{84}, 50001 (2008).

\bibitem{Kumar} A. Kumar, S. Adhikari, S. Banerjee, and S. Roy, Phys. Rev. A \textbf{87}, 022307 (2013).

\bibitem{ms} H. A. Carteret and A. Sudbery, J. Phys. A \textbf{33}, 4981 (2000).

\bibitem{tangle} V. Coffman, J. Kundu, and W. K. Wootters, Phys. Rev. A \textbf{61}, 052306 (2000).

\bibitem{nonlocality} S. Ghose, N. Sinclair, S. Debnath, P. Rungta, and R. Stock, Phys. Rev. Lett. \textbf{102}, 250404 (2009).

\bibitem{equatorial1} D. Bru$\beta$, M. Cinchetti, G. M. Mauro D'Ariano, and C. Macchiavello, Phys. Rev. A \textbf{62},012302 (2000).

\bibitem{equatorial2} H. Fan, K. Matsumoto, X. B. Wang, and M. Wadati, Phys. Rev. A \textbf{65},012304 (2001).

\bibitem{2/3} S. Massar and S. Popescu, Phys. Rev. Lett. \textbf{74}, 1259 (1995).

\bibitem{2/3+} S. Popescu, Phys. Rev. Lett. \textbf{72}, 797 (1994).

\bibitem{mt} T. A. Brun and O. Cohen, Phys. Lett. A \textbf{281}, 88 (2001).

\end{thebibliography}
\end{document}